\documentclass[eng]{ajceam-class}
\usepackage{amsmath,amsthm,amsfonts,amssymb,amscd}
\usepackage{siunitx}
\usepackage{lastpage}
\usepackage{enumerate}
\usepackage{fancyhdr}
\usepackage{mathrsfs}
\usepackage{xcolor}
\usepackage{graphicx}
\usepackage{listings}
\usepackage{hyperref}
\usepackage[utf8]{inputenc}
\DeclareUnicodeCharacter{2212}{-}
\DeclareUnicodeCharacter{394}{$\Delta$}
\title{High Throughput Screening of Ternary Nitrides with Convolutional Neural Networks.}

\author[*]{Antony A. Ayieko}
\author[*]{Michael O. Atambo}
\author[*]{George O. Amolo}

\affil[*]{Materials Modelling Group, School of Physics and Earth Science, Technical University of Kenya.}

\firstauthor{Ayieko}

\contactauthor{M.O. Atambo}           
\email{michael.atambo@tukenya.ac.ke}            

\thismonth{October}
\thisyear{2022}

\abstract{
The development of new materials is a core aspect of advancement in synthesis and application for industry. There is a vast number 
 of possible chemical permutations of the basic elements that can be explored to synthesize  materials that possess attractive catalytic, mechanical and electrical properties that may not be easily accessible to traditional experimental methods for various reasons, including cost and time considerations. Nitrides, as examples, require very stringent and precise conditions to successfully synthesize making their experimental exploration very slow. In this paper, we employ the use of machine learning algorithms to predict the bulk properties of Ternary Metal Nitrides (TMN), specifically their bulk modulus which is correlated with the hardness of the material. We were able to develop a consistent model with encouraging accuracy, that was able to predict the bulk moduli of materials that previously did not have computed values. The model was trained on $10^3$ ternary materials with known elastic properties and defined structures, and was able to predict the bulk modulus of $\thickapprox 1,000$ Ternary Metal Nitrides (TMNs) to $\thickapprox 80\%$ accuracy. This approach is orders of magnitude faster than the traditional computational approaches like density functional theory (DFT)\cite{dft-paper} which makes exploratory identification of materials with promising properties fast. We propose that such models be used to select interesting candidates for high throughput computation from first principles. 
}


\begin{document}

\maketitle
\printcontactdata

\section{Introduction}
\firstword{T}{he}
 discovery, characterization, and synthesis of new materials is a core function of material science, however it is difficult to conduct experiments over extended periods of time with a large number of possible synthesis candidates taking into account the issue of equipment and resources. Because of these limitations, important discoveries have mostly occurred by human intuition or by chance. Moving beyond this, it is now imperative that the systematic discovery of materials with required properties be possible, if not outright design of the same from scratch. 

A new age of  accurate first-principles approach was born with the development of techniques such as  density functional theory (DFT)\cite{dft-paper}, Monte Carlo simulations\cite{Monte-Carlo-Simulation} and molecular dynamics (MD)\cite{Molecular-Dynamics-Simulation}. These techniques allowed the researchers quickly and rapidly calculate and/or simulate materials properties and behavior before actually synthesizing the material  through the use of High Throughput Computing.

With the onset of data-driven research, the use of machine learning algorithms became viable due to a large amount of data available in the public domain. This led to using of algorithms such as Decision trees, Convolutional Neural Networks (CNNs)\cite{cnn-article}, Support Vector Machines (SVMs)\cite{svm-paper}, Recurrent Neural Networks (RNNs)\cite{rnn-paper}, and Transformers\cite{attention-is-all-you-need}. These algorithms allowed for  development of even faster exploration tools required for rapid discovery of new, stable and useful materials.

This work was inspired by the limited material property data on ternary materials which have been observed to have  applications in fields such as solid-state lighting, ceramic hard coatings, and permanent magnets. The synthesis of these TMNs, however, comes with its challenges. Despite so much potential to revolutionize industries, the TMNs space remains sparsely studied in the literature, having just about 700 known ternary metal nitrides categorized in the Materials Project Database\cite{materials-project} and also indexed at the Inorganic Crystal Structure Database (ICSD)\cite{ICSD}. This number is very small considering there are over 48,000 ternary materials (TM) listed on the materials project that are also indexed at the ICSD.

This low number of known TMNs is due to the fact that solid-state nitride materials are very hard to synthesize in the average laboratory. This fact is brought about by nitride synthesis requiring very stringent and hard to achieve conditions including the presence of an oxygen-rich environment and a water-free atmosphere. The nitrides also easily decompose at high temperatures and are thus very unstable in their molecular form. The $N_2$ molecule is also very stable. 

\begin{figure}[htp]
    \centering
    \includegraphics[scale=0.5]{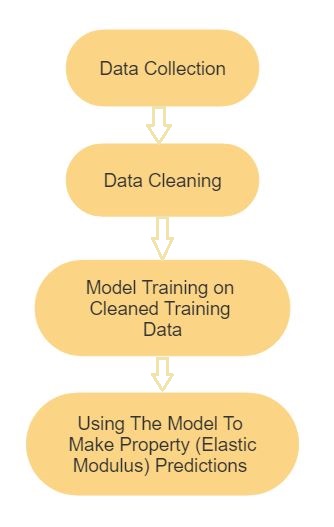}
    \caption{Flow chart showing the stages the study was carried out in.}
    \label{fig:workflow}
\end{figure}

\subsection{Ternary Metal Nitrides}
Ternary Metal nitrides are materials that have the chemical makeup of Metal-Metal-Nitrogen. They are composed of two metal elements and one nitrogen to collectively form a molecule with three elements.

\subsection{Machine Learning in Material Science}
Machine learning (ML)\cite{machine-learning-wikipedia}, a subset of artificial intelligence (AI)\cite{artificial-intelligence-wikipedia}, focuses on creating systems that learn from the data they consume. Artificial Intelligence can be used to describe any machine or system that imitates human intelligence. Machine Learning algorithms are becoming better over time for use in materials science and informatics. They have accelerated material discovery due to the ease of rapidly making property predictions for thousands of materials at a go using high-throughput methods as opposed to first principle methods.

In this study, we utilize a newly designed model that is a variation of the Convolutional Neural Networks (CNN)\cite{cnn-article} known as Crystal Graphic Convolutional Neural Network (CGCNN)\cite{PhysRevLett.120.145301}. This model encodes the crystal-graphic structure of a molecule into a Convolutional Neural Network Model in the form of a graph. This embeds both the atomic information and bonding interactions between atoms.

\subsection{Crystallographic representation for Neural Networks}
Crystal graphs represent the atomic structure of crystalline materials, where atoms are represented as nodes and bonds between atoms as edges. They capture the connectivity and spatial arrangement of atoms within a crystal lattice. Crystal Graph Convolutional Neural Networks (CGCNN) leverage these crystal graphs to analyze and predict material properties. 
\paragraph{}
Crystals are represented as graphs $G$ with nodes $vi$ for each atom $i$ and edges $uij$ for each bond between atoms $i$ and $j$. CGCNNs have convolutional layers that update each atom's features by combining it with its neighbors' features, capturing the local environment around each atom.

\begin{equation}
    v_i^{(t+1)} = f(v_i^t, v_j^t, u_{(i,j)}^k)
\end{equation}

Pooling layers summarize all the atom features into one vector representing the full crystal. 

\begin{equation}
    v_c = \text{Pool}(v_0, v_1,...,v_N)
\end{equation}

This crystal vector preserves important symmetries, including the ordering of atoms. 

Fully connected layers map the crystal vector to predict the target property like formation energy.
The model is trained on many example crystals with known properties from simulations. It learns to predict properties by minimizing errors.

\begin{figure}[htp]
    \centering
    \includegraphics[scale=0.5]{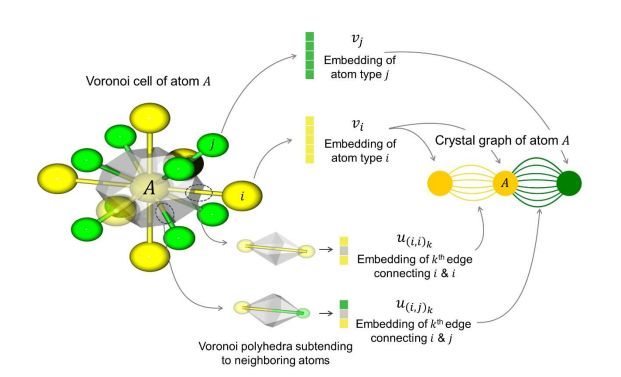}
    \caption{Illustration of iCGCNN crystal graph. The crystal graph shown on the far right represents the local environment of atom $A$. Multiple edges connect $A$ to neighboring nodes to show the number of Voronoi neighbors. The nodes and edges are embedded with vectors that characterize the constituent atoms ($v_i, v_j$) and their correlations with neighboring atoms (${U_{(i, j)}}_k, {U_{(i, j)}}_k$) respectively. Edge vectors include information about the Voronoi polyhedral such as solid angle, area, and volume. Image courtesy of APS \cite{icgcnn}}
    \label{fig:imporoved-cgcnn}
\end{figure}

CGCNNs apply convolutional neural network techniques specifically tailored to operate on crystal graphs, allowing them to learn and extract features from the atomic neighborhoods and capture complex relationships between the arrangement of atoms and material properties.
\paragraph{}
In the work of Xie and Grossman\cite{PhysRevLett.120.145301} they propose the use of a specialized variant of the Convolutional Neural network that encodes a molecule's crystalline structure in a graph and then converts the graph into a Neural Network that they named Crystal Graph Convolutional Neural Networks (CGCNN). This model presented significant reliability in the predicted mechanical properties of molecules due to the inclusion of the crystal structure in the data provided to the Neural Network. 
\paragraph{}
Park and Wolverton\cite{icgcnn} suggested an improvement on the CGCNN, which they called iCGCNN - Improved Crystal Graph Convolutional Neural Networks. The improved model is created by incorporating information on the Voronoi tessellated crystal structure as demonstrated in \textbf{Figure \ref{fig:imporoved-cgcnn}}, explicit 3-body correlations of neighboring constituent atoms, and an optimized chemical representation of inter-atomic bonds in the crystal graphs. Using both CGCNN and iCGCNN, they were able to  screen $10^5$ compounds, and accelerating the computational time of the high-throughput search by a factor of 130.

\paragraph{}
\textbf{Bartel et al.} \cite{TMNs} was able to demonstrate the use of machine learning and computational techniques in the discovery of new ternary materials. They collected data from the ICSD Database and trained a data-mined structure predictor (DMSP)\cite{dmsp-paper} algorithm that helped predict stable crystal structures, the stable structures were then narrowed down to a hand full that were then accurately calculated using first principle i.e DFT. They were able to narrow down and experimentally learn of 7 new TMNs using this method.

\section{Methodology}
In this section, we will discuss the method used to process the data obtained from the Materials Project database\cite{materials-project}, the performance of the model, the selection of a data size for training and testing and also the performance of the model on newly presented data (materials).

\subsubsection{Data}
We were able to collect data for more than 59,000 materials which included the Crystallographic Information (CIF formatted), the Chemical Formula, the material's electronic band gap, the Elastic Properties[Bulk Modulus, Shear Modulus] and finally the materials ICSD indexing id.

However there were only a few materials that though they had CIF\cite{cif-file-intro} file information available, we determined to be unusable since they demonstrated negative values for the predicted bulk modulus, raising questions about their structures.

We earlier discussed that we decided to use all ternary materials for the training of the model. This was because the small data-set that comes with strictly using data from TMNs. We ran the learning model on all the data-sets and recorded the Mean Absolute Error (MAE)\cite{mae-explanation} of each model with respect to the dataset size as seen in Table \ref{table:dataset-size-perfomance}.

\subsubsection{iGCNN Model}
We used a modified version of the CGCNN model \cite{PhysRevLett.120.145301} with inspiration from the improved version of the CGCNN \cite{icgcnn} by using additional descriptors for the model extracted from the crystal graphs which include the information of the Voronoi tessellated crystal structure \cite{voronoi-tessellations}, explicit three-body correlations of neighboring constituent atoms, and an optimized chemical representation of inter-atomic bonds, all of which are absent in the original crystal graph models utilized by the original CGCNN framework \cite{PhysRevLett.120.145301}

\section{Results and Discussion}
\subsubsection{Data Dependence}
For various amounts of data, a model was generated and its accuracy tested, because of the limited availability of data, going beyond the limited numbers available in open databases required introducing classes of closely related materials. Thus, the data
dependence test was done against  Ternary Metal Nitrides, Ternary Nitrides,  Ternary Metals and Ternary Materials, in increasing levels of number of samples available for training. The results are shown in Tab.~\ref{table:dataset-size-perfomance}, showing an increase in accuracy from 37\% MAE to 18\% MAE.
{\centering
    \begin{table}
    \begin{tabular}{c  c c } 
        \hline\hline
        \textbf{Query type} & \textbf{Number of materials} & \textbf{MAE}   \\ [0.5ex] 
        \hline\hline
        Ternary Materials & 59,023 & 16.76       \\  [1ex]
        
        Ternary Metals & 42,042  &  18.53        \\ [1ex]
        
        Ternary Nitrides & 4,529  & 31.23        \\ [1ex]
        
        Ternary Metal Nitrides & 3,153  & 37.15   \\ [1ex]
        \hline\hline
    \end{tabular}
    \caption{
        \label{table:dataset-size-perfomance}
        Table showing the performance of the model when fed varying amounts of training data. It can clearly be observed that a larger dataset reduces the MAE to acceptable ranges.
    }
    \end{table}
}

\subsubsection{Model Performance in Training}
\begin{figure}[htp]
    \centering
    \includegraphics[scale=0.6]{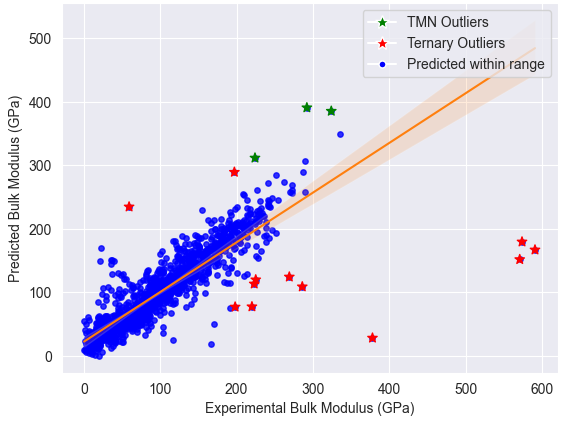}
    \caption{A regression scatter plot showing the relationship between the predicted and actual bulk modulus values after training. A total of  1474 points are shown, and  highlighted are the outliers with poorly predicted bulk-modulus during, these are not necessarily TMN's.}
    \label{fig:scatter-plt-outliers}
\end{figure}

The data we collected on the material properties of the TMNs was spilt in the ratio; 6:2:2 for training, validation, and testing respectively. We found this ratio to be the best as most of the data is fed to the model for training and the remaining is split equally for validation and testing. 
The data we collected on the material properties of the TMNs was spilt in the ratio; 6:2:2 for training, validation, and testing respectively. We found this ratio to be the best as most of the data is fed to the model for training and the remaining is split equally for validation and testing.  Fig.~\ref{fig:scatter-plt-outliers} shows a scatter plot with a regression line  demonstrating  how well the model performed on the test data.
On the independent axis is the experimental data, and the dependent axis the predicted value, the clustering around the regression line indicates agreement between the two.

\subsubsection{Making Predictions}
\begin{figure}[htp]
    
    \includegraphics[width=9.3cm,height=7cm]{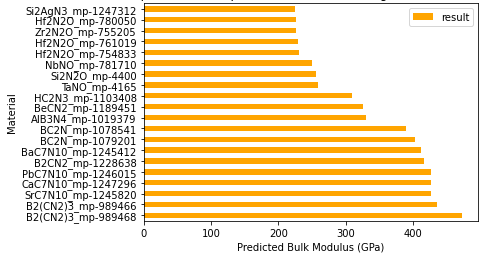}
    \caption{A bar graph showing the predicted value of the bulk modulus of a selection of the TMNs without calculated bulk modulus in the Materials Project Database\cite{materials-project}.}
    \label{fig:bar-predict}
\end{figure}

After training the model, we then saved the best-performing version of the model and tested it against data that did not have any calculated elastic property. This set was taken from the larger dataset of all the collected ternary materials and was filtered out using the logic of band-gap to get the metals and logic of the Metal-Metal-Nitrogen Makeup to get only TMNs. This subset was then further filtered to get TMNs \textit{without} known bulk moduli. This final subset was then fed to the model and the prediction script was run. Fig.~\ref{fig:bar-predict} shows the predicted value of the bulk modulus for the subset of 50 candidates with the highest predicted bulk modulus.

Classes of materials such as the B-C-N system and ternary metal nitrides that contain Barium, Boron or Hafnium show great promise in producing materials that are super-hard. Some of the materials in these classes have been studied showing their hardness which is directly linked to their high bulk modulus value. In the B-C-N class, B2(CN2)3, B2CN2, BaC7N10 and BC2N where predicted to be within the top ten TMN materials with the highest bulk modulus. These predictions are in line with investigations of the mechanical and electrical properties \cite{dr_mike_bcn} of BC2N and c-BN using accurate Quantum Monte Carlo\cite{Monte-Carlo-Simulation} techniques. We observed generally that TMNs that contained \textbf{Barium, Boron or Hafnium} in their make-up demonstrated a very high predicted bulk modulus.

\subsubsection{Model Performance For Non TMNs}
\begin{figure}[htp]
    \centering
    \includegraphics[scale=0.6]{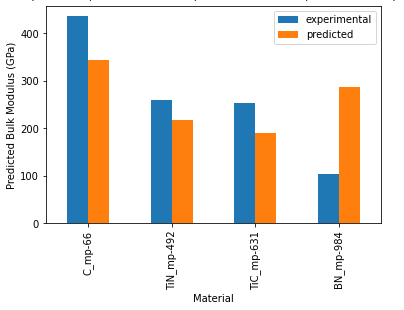}
    \caption{A bar graph showing the predicted value of the bulk modulus of common Hard Materials such as Diamond, Titanium Carbide, Titanium nitride, and one outlier,  Boron Nitride in GPa.}
    \label{fig:bar-predict-common}
\end{figure}

We used the model to test its accuracy in predicting the bulk modulus of commonly known materials such as Diamond, Titanium Carbide, Titanium Nitride, and Boron Nitride. On running the prediction, the results obtained are shown in Figure \ref{fig:bar-predict-common}. The agreement is stronger for materials with metals, indicating the models input data that was  biased towards metal nitrides.

\section{Conclusions}

Our results reinforce the utility of employing Machine Learning in materials science and technology for the purposes of  prediction of the bulk properties of a large number of materials of interest with relatively low computational effort and time requirements compared to traditional electronic structure methods.  It is observed that up to 70 - 80 percent of the materials under study have results comparable to those obtained from first principles calculations. 
We have also been able to predict the bulk modulus of several theoretical structures, that have not been previously studied, that show promising hardness, appropriate for hard materials applications.

\section{Recommendations}
Our model can be used to filter out outliers in materials databases, such as those with predicted negative values of bulk moduli (indicating incorrect structures deposited in the database), to improve the quality of the data stored therein, and correct obvious anomalies in the available data. The theoretical structures containing nitrides  predicted to have high bulk moduli are interesting candidates for synthesis and characterization.

\section{Acknowledgments}

We would like to acknowledge KENET and CHPC (project MATS862). A.A. acknowledges the useful discussions with Gerald Okioma of the Materials Science Group, Technical University of Kenya. 

\insertbibliography{References}

\begin{thebibliography}{10}
\providecommand{\url}[1]{#1}
\csname url@samestyle\endcsname
\providecommand{\newblock}{\relax}
\providecommand{\bibinfo}[2]{#2}
\providecommand{\BIBentrySTDinterwordspacing}{\spaceskip=0pt\relax}
\providecommand{\BIBentryALTinterwordstretchfactor}{4}
\providecommand{\BIBentryALTinterwordspacing}{\spaceskip=\fontdimen2\font plus
\BIBentryALTinterwordstretchfactor\fontdimen3\font minus
  \fontdimen4\font\relax}
\providecommand{\BIBforeignlanguage}[2]{{%
\expandafter\ifx\csname l@#1\endcsname\relax
\typeout{** WARNING: IEEEtran.bst: No hyphenation pattern has been}%
\typeout{** loaded for the language `#1'. Using the pattern for}%
\typeout{** the default language instead.}%
\else
\language=\csname l@#1\endcsname
\fi
#2}}
\providecommand{\BIBdecl}{\relax}
\BIBdecl

\bibitem{dft-paper}
\BIBentryALTinterwordspacing
W.~Kohn, A.~D. Becke, and R.~G. Parr, ``Density functional theory of electronic
  structure,'' \emph{The Journal of Physical Chemistry}, vol. 100, no.~31, pp.
  12\,974--12\,980, Jan. 1996. [Online]. Available:
  \url{https://doi.org/10.1021/jp960669l}
\BIBentrySTDinterwordspacing

\bibitem{Monte-Carlo-Simulation}
\BIBentryALTinterwordspacing
P.~L. Bonate, ``A brief introduction to monte carlo simulation,''
  \emph{Clinical Pharmacokinetics}, vol.~40, no.~1, pp. 15--22, 2001. [Online].
  Available: \url{https://doi.org/10.2165%2F00003088-200140010-00002}
\BIBentrySTDinterwordspacing

\bibitem{Molecular-Dynamics-Simulation}
\BIBentryALTinterwordspacing
S.~A. Hollingsworth and R.~O. Dror, ``Molecular dynamics simulation for all,''
  \emph{Neuron}, vol.~99, no.~6, pp. 1129--1143, Sep. 2018. [Online].
  Available: \url{https://doi.org/10.1016/j.neuron.2018.08.011}
\BIBentrySTDinterwordspacing

\bibitem{cnn-article}
\BIBentryALTinterwordspacing
K.~O'Shea and R.~Nash, ``An introduction to convolutional neural networks,''
  2015. [Online]. Available: \url{https://arxiv.org/abs/1511.08458}
\BIBentrySTDinterwordspacing

\bibitem{svm-paper}
\BIBentryALTinterwordspacing
M.~Hearst, S.~Dumais, E.~Osuna, J.~Platt, and B.~Scholkopf, ``Support vector
  machines,'' \emph{{IEEE} Intelligent Systems and their Applications},
  vol.~13, no.~4, pp. 18--28, jul 1998. [Online]. Available:
  \url{https://doi.org/10.1109%2F5254.708428}
\BIBentrySTDinterwordspacing

\bibitem{rnn-paper}
\BIBentryALTinterwordspacing
A.~Sherstinsky, ``Fundamentals of recurrent neural network (rnn) and long
  short-term memory (lstm) network,'' 2018. [Online]. Available:
  \url{https://arxiv.org/abs/1808.03314}
\BIBentrySTDinterwordspacing

\bibitem{attention-is-all-you-need}
\BIBentryALTinterwordspacing
A.~Vaswani, N.~Shazeer, N.~Parmar, J.~Uszkoreit, L.~Jones, A.~N. Gomez,
  L.~Kaiser, and I.~Polosukhin, ``Attention is all you need,'' 2017. [Online].
  Available: \url{https://arxiv.org/abs/1706.03762}
\BIBentrySTDinterwordspacing

\bibitem{materials-project}
A.~Jain, S.~P. Ong, G.~Hautier, W.~Chen, W.~D. Richards, S.~Dacek, S.~Cholia,
  D.~Gunter, D.~Skinner, G.~Ceder, and K.~A. Persson, ``Commentary: The
  materials project: A materials genome approach to accelerating materials
  innovation,'' \emph{APL Mater.}, vol.~1, no.~1, p. 011002, 2013.

\bibitem{ICSD}
F.~Karlsruhe. (2022) Inorganic crystal structure database.
  \url{https://icsd.products.fiz-karlsruhe.de/en/about/about-icsd}
  (16/10/2022).

\bibitem{machine-learning-wikipedia}
{Wikipedia contributors}, ``Machine learning,''
  \url{https://en.wikipedia.org/w/index.php?title=Machine_learning&oldid=1122823523},
  Nov. 2022, accessed: NA-NA-NA.

\bibitem{artificial-intelligence-wikipedia}
------, ``Artificial intelligence,''
  \url{https://en.wikipedia.org/w/index.php?title=Artificial_intelligence&oldid=1123242409},
  Nov. 2022, accessed: NA-NA-NA.

\bibitem{PhysRevLett.120.145301}
\BIBentryALTinterwordspacing
T.~Xie and J.~C. Grossman, ``Crystal graph convolutional neural networks for an
  accurate and interpretable prediction of material properties,'' \emph{Phys.
  Rev. Lett.}, vol. 120, p. 145301, Apr 2018. [Online]. Available:
  \url{https://link.aps.org/doi/10.1103/PhysRevLett.120.145301}
\BIBentrySTDinterwordspacing

\bibitem{icgcnn}
\BIBentryALTinterwordspacing
C.~W. Park and C.~Wolverton, ``Developing an improved crystal graph
  convolutional neural network framework for accelerated materials discovery,''
  \emph{Physical Review Materials}, vol.~4, no.~6, Jun. 2020. [Online].
  Available: \url{https://doi.org/10.1103/physrevmaterials.4.063801}
\BIBentrySTDinterwordspacing

\bibitem{TMNs}
\BIBentryALTinterwordspacing
W.~Sun, C.~J. Bartel, E.~Arca, S.~R. Bauers, B.~Matthews, B.~Orva{\~{n}}anos,
  B.-R. Chen, M.~F. Toney, L.~T. Schelhas, W.~Tumas, J.~Tate, A.~Zakutayev,
  S.~Lany, A.~M. Holder, and G.~Ceder, ``A map of the inorganic ternary metal
  nitrides,'' \emph{Nature Materials}, vol.~18, no.~7, pp. 732--739, Jun. 2019.
  [Online]. Available: \url{https://doi.org/10.1038/s41563-019-0396-2}
\BIBentrySTDinterwordspacing

\bibitem{dmsp-paper}
\BIBentryALTinterwordspacing
G.~Hautier, C.~Fischer, V.~Ehrlacher, A.~Jain, and G.~Ceder, ``Data mined ionic
  substitutions for the discovery of new compounds,'' \emph{Inorganic
  Chemistry}, vol.~50, no.~2, pp. 656--663, dec 2010. [Online]. Available:
  \url{https://doi.org/10.1021%2Fic102031h}
\BIBentrySTDinterwordspacing

\bibitem{cif-file-intro}
\BIBentryALTinterwordspacing
S.~R. Hall, J.~D. Westbrook, N.~Spadaccini, I.~D. Brown, H.~J. Bernstein, and
  B.~McMahon, ``Specification of the crystallographic information file
  ({CIF}),'' in \emph{International Tables for Crystallography}.\hskip 1em plus
  0.5em minus 0.4em\relax International Union of Crystallography, Oct. 2006,
  pp. 20--36. [Online]. Available:
  \url{https://doi.org/10.1107/97809553602060000728}
\BIBentrySTDinterwordspacing

\bibitem{mae-explanation}
\BIBentryALTinterwordspacing
W.~Wang and Y.~Lu, ``Analysis of the mean absolute error ({MAE}) and the root
  mean square error ({RMSE}) in assessing rounding model,'' \emph{{IOP}
  Conference Series: Materials Science and Engineering}, vol. 324, p. 012049,
  Mar. 2018. [Online]. Available:
  \url{https://doi.org/10.1088/1757-899x/324/1/012049}
\BIBentrySTDinterwordspacing

\bibitem{voronoi-tessellations}
\BIBentryALTinterwordspacing
L.~Ward, R.~Liu, A.~Krishna, V.~I. Hegde, A.~Agrawal, A.~Choudhary, and
  C.~Wolverton, ``Including crystal structure attributes in machine learning
  models of formation energies via voronoi tessellations,'' \emph{Physical
  Review B}, vol.~96, no.~2, Jul. 2017. [Online]. Available:
  \url{https://doi.org/10.1103/physrevb.96.024104}
\BIBentrySTDinterwordspacing

\bibitem{dr_mike_bcn}
\BIBentryALTinterwordspacing
M.~O. Atambo, ``Mechanical and electronic properties of cubic boron nitride and
  carbon ...'' 2013. [Online]. Available:
  \url{http://41.89.164.27:8080/xmlui/handle/123456789/982}
\BIBentrySTDinterwordspacing

\end{thebibliography}
\end{document}